\documentclass[12pt,superscriptaddress]{revtex4}
\usepackage{graphicx}
\begin{document}

\title{Clarifications to Questions and Criticisms \\on the Johansen-Ledoit-Sornette Financial Bubble Model}

\author{Didier Sornette}
 \email{dsornette@ethz.ch}
 \affiliation{Department of Management, Technology and Economics, ETH Zurich, Kreuzplatz 5, CH-8032 Zurich, Switzerland} %
 \affiliation{Swiss Finance Institute, c/o University of Geneva, 40 blvd. Du Pont d'Arve, CH-1211 Geneva 4, Switzerland}%

\author{Ryan Woodard}
 \email{rwoodard@ethz.ch}
 \affiliation{Department of Management, Technology and Economics, ETH Zurich, Kreuzplatz 5, CH-8032 Zurich, Switzerland} %

\author{Wanfeng Yan}
 \email{wyan@ethz.ch}
 \affiliation{Department of Management, Technology and Economics, ETH Zurich, Kreuzplatz 5, CH-8032 Zurich, Switzerland} %

\author{Wei-Xing Zhou}
 \email{wxzhou@ecust.edu.cn}
 \affiliation{School of Business, East China University of Science and Technology, Shanghai 200237,
 China}

\begin{abstract}
The Johansen-Ledoit-Sornette (JLS) model of rational expectation bubbles with
finite-time singular crash hazard rates has been developed to describe the
dynamics of financial bubbles and crashes. It has been applied successfully to
a large variety of financial bubbles in many different markets. Having been
developed over a decade ago, the JLS model has been studied, analyzed, used and
criticized by several researchers. Much of this discussion is helpful for
advancing the research. However, several serious misconceptions seem to be
present within this literature both on theoretical and empirical aspects.
Several of these problems stem from the fast evolution of the literature on the
JLS model and related works. In the hope of removing possible misunderstanding
and of catalyzing useful future developments, we summarize these common
questions and criticisms concerning the JLS model and synthesize the current
state of the art and existing best practice.


\end{abstract}

\maketitle \clearpage

\section{Introduction}

The Johansen-Ledoit-Sornette (JLS) model \cite{js,jsl,jls,sornettecrash} has
been developed to describe the dynamics of financial bubbles and crashes. The
model states that bubbles are not characterized by an exponential increase of
price but rather by a faster-than-exponential growth of price. This phenomenon is
generated by behaviors of investors and traders that create positive feedback
in the valuation of assets leading to unsustainable growth ending with a finite-time
singularity at some future time $t_c$.

One can identify two broad classes of positive feedback mechanisms.
The first technical class includes (i) option hedging \cite{SircarPapa}, (ii) insurance
portfolio strategies (see
paragraph 1 on page 380 of Ref.\cite{Longshleifersummers90}
stating that ``Positive feedback trading is also exhibited by buyers of portfolio insurance...'')
and Refs.\cite{Black88,KALIFE2004}),
(iii) market makers bid-ask spread in response to past
volatility \cite{Pasquariello2000,Angeletal2010},
(iv) learning of business networks and human capital build-up \cite{Mithas08,Gislersorwood11}, (v)
procyclical financing of firms by banks (boom vs contracting times) \cite{Li-Erica}, (vi) trend
following investment strategies, (vii) asymmetric information on hedging
strategies \cite{FenderIngo} (viii) the interplay of mark-to-market accounting and regulatory
capital requirements \cite{AtikJeffery,AllenLinda}. The second class
of positive feedback mechanisms is behavioral
and emphasizes that positive feedbacks emerge
as a result of the propensity of humans to imitate, of their social gregariousness
and the resulting herding. This critical time $t_c$ of the model is interpreted
as the end of the bubble, which is often but not necessarily the time when a
crash occurs in the actual system. During this growth phase, the tension and
competition between the value investors and the noise traders create deviations
around the hyperbolic power law growth in the form of oscillations that are periodic in
the logarithm of the time to $t_c$. Combining these two effects, this model
succinctly describes the price during a bubble phase as a log-periodic
(hyperbolic) power law (LPPL).

Since its introduction, the JLS model has been used widely to detect
bubbles and crashes ex-ante (i.e., with advanced documented notice in real
time) in various kinds of markets such as the 2006-2008 oil bubble \cite{oil},
the Chinese index bubble in 2009 \cite{Jiangetal09}, the real estate market in
Las Vegas \cite{ZhouSorrealest08}, the U.K.~and U.S.~real estate bubbles
\cite{Zhou-Sornette-2003a-PA,Zhou-Sornette-2006b-PA,fry,Fry-2010a,Fry-2010b}, the
Nikkei index anti-bubble in 1990-1998 \cite{JohansenSorJapan99} and the S\&P
500 index anti-bubble in 2000-2003 \cite{SorZhoudeeper02}. Other recent
studies performed in an ex-post mode include the Dow Jones Industrial Average historical bubbles
\cite{Vandewalle-Ausloos-Boveroux-Minguet-1999-EPJB}, the corporate bond
spreads \cite{Clark-2004-PA}, a Polish stock market bubble
\cite{Gnacinski-Makowiec-2004-PA}, some western stock market bubbles
\cite{Bartolozzi-Drozdz-Leinweber-Speth-Thomas-2005-IJMPC}, U.K.~stock market
bubbles \cite{Fry-2012}, the Brazilian real (R\$) - US dollar (USD) exchange
rate \cite{Matsushita-daSilva-Figueiredo-Gleria-2006-PA}, 2000-2010 world major
stock indices \cite{Drozdz-Kwapien-Oswiecimka-Speth-2008-APPA}, South African
stock market bubbles \cite{ZhouSorrSA09}, the US repurchase agreements market
\cite{repo} and emerging markets \cite{js2001,Fry-Masood-2011}. Moreover, new
experiments in ex-ante bubble detection and forecasts have been launched since
November 2009 in the Financial Crisis Observatory at ETH Zurich
\cite{BFE-FCO09, BFE-FCO10, BFE-FCO11}.
In many cases, market risk is contagious \cite{Malsornbook2005} and
market corrections and crashes occur successively in
a short period \cite{zhousorworld03} including the latest financial crisis, which is partially due to
globalization and assets diversification such that the US stock market and
world financial stock market exhibit long-range cross-correlations \cite{Podobnik2010,Wang2011}.

Having been developed over a decade ago, the JLS model has been studied, used
and criticized by many researchers including Feigenbaum \cite{feigenbaum},
Chang and Feigenbaum \cite{Changfeigenbaum06,Changfeigenbaum08}, van Bothmer
and Meister \cite{bm}, Fry \cite{fry}, and Fantazzini and Geraskin \cite{fg}.
The most recent papers addressing the pros and cons of past works on the JLS
model are written by Bree and his collaborators \cite{bree1, bree2}. Many ideas
in these last two papers are correct, pointing out some of the
inconsistencies in earlier works. However, there are some serious
misunderstandings concerning both the theoretical and empirical parts of the
model. Therefore, it is necessary to address and clarify the misconceptions
that some researchers seem to hold concerning the JLS model and to provide an
updated, concise reference on the JLS model.

The structure of this paper is as follows. Section \ref{sec:theory} discusses
the questions about the theory and derivation of the JLS model. The questions
on fitting methods of the model are commented on Section \ref{sec:fitting}.
Issues on probabilistic forecast will be addressed in Section
\ref{sec:predict}. We conclude in Section \ref{sec:conclusion}.

\section{Presentation and remarks on the theoretical foundation of the JLS model}
\label{sec:theory}

We will give the derivation of the JLS model first in this section.
Then we discuss three issues related to the derivation and the
proper parameter ranges.

\subsection{Derivation of the JLS model}
\label{sec:jlsderive}

The JLS model starts from the rational expectation settings of
\cite{bwrationalexpectation}, where the observed price $p_o$ of a stock can be written as
\begin{equation}
p_o = p^* + p~,
\label{eq:bw}
\end{equation}
where $p^*$ and $p$ represent respectively the fundamental value and the bubble component.
Eq.~(\ref{eq:bw}) shows that the price is a linear combination of
the fundamental value and the bubble component. The JLS model specifies the
dynamics of the bubble component \emph{independently} of the dynamics of the
fundamental price. The later can be specified according to standard valuation models,
for instance leading to the usual geometrical random walk benchmark.
The JLS model adds to this featureless fundamental price
the so-called log-periodic power law structure, which is used to diagnose
the presence of bubbles. Lin et al. \cite{lin} have considered a self-consistent
mean-reverting process for $p^*$ that makes consistent the calibration of the observed
price $p_o$ by the JLS model.

The JLS model starts from the assumption that
the dynamics of the bubble component of the price satisfies a simple stochastic
differential equation with drift and jump:
\begin{equation}
\frac{dp}{p} = \mu(t) dt + \sigma dW - \kappa dj, \label{eq:jlsdynamics}
\end{equation}
where $p$ is the stock market bubble price, $\mu(t)$ is the drift (or trend) and $dW$
is the increment of a Wiener process (with zero mean and unit variance). The
term $dj$ represents a discontinuous jump such that $j = 0$ before the crash
and $j = 1$ after the crash occurs. The loss amplitude associated with the
occurrence of a crash is determined by the parameter $\kappa$. Each successive
crash corresponds to a jump of $j$ by one unit. The dynamics of the jumps is
governed by a crash hazard rate $h(t)$. Since $h(t)dt$ is the probability that
the crash occurs between $t$ and $t + dt$ conditional on the fact that it has
not yet happened, we have $E_t[dj] = 1 \times h(t)dt + 0 \times (1 - h(t)dt)$
and therefore the expectation of $dj$ is given by
\begin{equation}
E_t[dj] = h(t)dt.
\label{eq:jlsjump}
\end{equation}

Under the assumption of the JLS model, noise traders exhibit collective herding
behaviors that may destabilize the market. The model assumes that the aggregate
effect of noise traders can be accounted for by the following dynamics of the
crash hazard rate:
\begin{equation}
h(t) = B'(t_c - t)^{m-1} + C'(t_c - t)^{m-1} \cos(\omega \ln(t_c -
t)-\phi').
\label{eq:hazard}
\end{equation}
The cosine part of the second term in the r.h.s. of
(\ref{eq:hazard}) takes into account the existence of possible
hierarchical cascades \cite{SornetteJohansen97}
of accelerating panic punctuating the growth
of the bubble, resulting from a preexisting hierarchy in noise
trader sizes \cite{zhoudunbar05} and/or the interplay between market price impact
inertia and nonlinear fundamental value investing \cite{IdeSornette}.
Expression (\ref{eq:hazard}) also contains a hyperbolic power law growth ending
at a finite-time singularity, which embodies the positive feedbacks
resulting from the technical and behavioral mechanisms summarized
above in the Introduction.

The non-arbitrage
condition expresses that the unconditional expectation $E_t[dp]$
of the price increment must be $0$, which leads to
\begin{equation}
\mu(t) \equiv {\rm E}\left[{dp/dt \over p}\right]_{\rm no~ crash} = \kappa h(t)~,
\label{hyjuyjuy5}
\end{equation}
by taking the expectation of (\ref{eq:jlsdynamics}).  Note that $\mu(t)dt$
is the return ${dp \over p}$ over the infinitesimal time interval $dt$
in the absence of crash. Using this and
substituting (\ref{eq:hazard}) and integrating yields the so-called
log-periodic power law (LPPL) equation:
\begin{equation}
\ln E[p(t)] = A + B(t_c - t)^m + C(t_c - t)^m \cos(\omega \ln(t_c -
t) - \phi)
\label{eq:lppl}
\end{equation}
where $B = -\kappa B'/m$ and $C = -\kappa C'/\sqrt{m^2+\omega^2}$. Note that
this expression (4) describes the average price dynamics only up to the end of
the bubble. The JLS model does not specify what happens beyond $t_c$. This
critical time $t_c$ is the termination of the bubble regime and the transition
time to another regime. The parameter $t_c$ represents the non-random
time of the termination of the bubble. However, its precise value is not known
with absolute precision, and its estimation can be written as
\begin{equation}
t_c^{estimated} = t_c^{true} + \epsilon~,
\end{equation}
where $\epsilon$ is an error term distributed according to some
distribution, while $t_c^{true}$ is deterministic.
Lin and Sornette \cite{LinSornette09} have recently extended
the modeling to include a stochastic mean reversal dynamics of the
critical time $t_c$, thus capturing the uncertain anticipation of
investors concerning the end of the bubble.

\subsection{Why $m$ should be between 0 and 1? A balancing act}

The mechanism of positive feedback leading to faster-than-exponential price
growth is captured by the exponent $m$, which should fall in the range $[0,
1]$. A negative value of $m$ would correspond to unrealistic diverging prices
in finite time. A value of $m$ larger than $1$ would correspond to a
decelerating price.

1. For $m < 1$, the crash hazard rate accelerates up to $t_c$ but
its integral up to $t$, which controls the total probability for a
crash to occur up to $t$, remains finite and less than 1 for all
times $t \leq t_c$. It is this property that makes it rational for
investors to remain invested, knowing that a bubble is developing and
that a crash is looming. Indeed, there is still a finite probability
that no crash will occur during the lifetime of the bubble and beyond,
so there is a chance for investors to gain from the ramp-up
of the bubble and walk away unscathed. The
excess return $\mu(t) = \kappa h(t)$ is the remuneration that
investors require to remain invested in the bubbly asset, which is
exposed to a crash risk. The crash hazard may diverge as $t$
approaches a critical time $t_c$, corresponding to the end of the
bubble.

2. Within the JLS framework, a bubble exists when the crash hazard
rate accelerates with time. According to (\ref{eq:hazard}), this
imposes $m < 1$ and $B' > 0$. That is, $m \geq 1$ cannot lead to an
accelerating hazard rate.

3. Finally, the condition that the price remains finite at all time,
including $t_c$, imposes that $m > 0$.

Bree \& Joseph \cite{bree1} note astutely that
imposing the constraint $m<1$ might be detrimental to the discovery
of the best fit of the LPPL model to a given price time series. Specifically,
they write: ``We presume that the reason that any fit with was rejected is because
then the increase in the index is exponentially declining whereas the
underlying mechanism requires it to be increasing. An alternative technique
would have been to place no restriction on the value of $m$, and if a value of
$m> 1$ is found, to reject the model, as we have done for the requirement that
the fitted LPPL never decreases.''

In other words, Bree \& Joseph \cite{bree1} propose to relax the condition
$0 < m < 1$ because the existence of best fits that have exponents outside this range
can be used as a diagnostic that no JLS-type bubble is present.
There is merit in this reasoning but one must also take into
account the large noise amplitude of the empirical price time series
together with the nonlinear nature of the LPPL model, which make
not fully reliable the
selection of best fits solely based on the minimization of the residual
root-mean-square errors (RMSE). We suggest that the use of constraints is to impose
that the statistical calibration exercises pass the financial conditions
of good sense (or ``smell test'', to paraphrase the term used by Robert Solow
in his testimony to the US House Committee on Science and Technology,
Subcommittee on Investigations and Oversight in July 2010 in the aftermath
of the financial crisis).
Other methods can be developed to address the compromise between
best fit and meaningfulness, by addressing directly the problem of noise
using bootstrap and ensemble methods, some of which having been
already implemented in the advance tests of the ETH Zurich financial crisis observatory
\cite{BFE-FCO09, BFE-FCO10, BFE-FCO11}.

Finally, this condition $0 < m < 1$ should not be understood
as ``protecting'' the JLS model, i.e., forcing the formula to fit the data.
Indeed, one should not confuse the fact of restricting the range of
$m$ values in the fitting procedure with the process of
qualifying or not the existence of a bubble. In the many studies
mentioned above, other selection procedures have been used
than based on the value of the exponent $m$.

\subsection{Non-negative risk condition}
\label{sec:positiverisk}

van Bothmer and Meister derived a constraint on the variables of the JLS model
\cite{bm} from the statement that the crash rate should be non-negative. It
states that:
\begin{equation}
b := -Bm - |C|\sqrt{m^2 + \omega^2} \geq 0. \label{eq:positiverisk}
\end{equation}
Most current research using the JLS model has taken this restriction into
consideration. It is among the basic restrictive filters for identifying
bubbles in a more modern framework. In
\cite{SorZhouforecast06,rebound,phpro,YanRebWooSor},
the parameter $b$ in
(\ref{eq:positiverisk}) is even used as a key trait implemented
within a pattern recognition
method to detect market rebounds.

\subsection{Decreasing price and decreasing expected price in the JLS model}

The condition derived by Graf v. Bothmer and Meister \cite{bm}
is equivalent to the statement that the expected price variation in the absence of
crash is non-negative, as seen from expression (\ref{hyjuyjuy5}).
This equivalence is emphasized by Bree and Joseph \cite{bree1},
who propose to use this requirement as a testable prediction.
And they note that many of the case studies reported in Johansen and Sornette \cite{js2001}
and in Sornette and Johansen \cite{SorJohansenQF01} have their
LPPL fits that exhibit a negative slope some of the time. Since the LPPL fit
is assumed to capture the expected price dynamics, a negative slope of the price
at certain times is in contradiction with expression (\ref{hyjuyjuy5}), given that the
crash hazard rate $h(t)$ is non-negative.
Bree and Joseph \cite{bree1} then conclude that this
fact is sufficient to reject the martingale condition
as being the mechanism underlying the LPPL to pre-crash bubbles, at least
for a large fraction of them. Their point is well taken and convincing.
However, another viewpoint is that condition (\ref{eq:positiverisk})
associated with the non-negativity of $h(t)$ and $\mu(t)$ via
the martingale condition offers a classification of bubbles in different types,
those that obey the martingale condition and those that do not.
Indeed, we do not believe that all bubbles are the same. For instance, we have
reported evidence in the past that some bubbles are ``fearful'' while
others are ``fearless''  \cite{AndersenSor04}.
 Moreover, one can imagine other ways to cure the
 possible occurrence of negative values of the parameter
 $b$ given by expression (\ref{eq:positiverisk}), which considers hierarchies of
 crashes, without the need to abandon the martingale condition.

We stress that this should not be confused with a more naive view that, during a bubble
following the JLS process, price has to always be increasing. Indeed,
the definition of the JLS model includes implicitly the stochastic
term $\sigma dW$ as in expression (\ref{eq:jlsdynamics}). In expectations, this
term disappears as does any unbiased random walk whose average position
remains identically zero while exhibiting realizations with significant deviations,
hence it is not included in the description of the initial JLS
paper. In the presence of this stochastic term $\sigma dW$, the price
can exhibit transient negative spells, even if the expected trend should be positive
according to expression (\ref{hyjuyjuy5}). This formulation is
nothing but that of the rational expectation of Blanchard and Watson
\cite{bwrationalexpectation},
which follows exactly the same procedure, with a stochastic component which
does not play a role in the specification of the crash hazard rate relationship
to the $\mu$ term, but is present to ensure that the
realized price can indeed decrease transiently, while the price should only
increase in expectations.

\subsection{Faster-than-exponential growth in the JLS model}

One of the fundamental differences between the JLS model and standard models of
financial bubbles is that the JLS model claims that the price follows a
faster-than-exponential growth rate during the bubble. It is necessary to
emphasize this statement as many researchers make mistakes here. For example,
Bree and Joseph wrote ``exponential growth is posited in the LPPL'' in several
places in \cite{bree1}.

Financial bubbles are generally defined as a transient upward acceleration of
prices above the fundamental value \cite{Galbraith,kindleberger,sornette2003}.
However, identifying unambiguously the presence of a bubble remains an unsolved
problem in standard econometric and financial economic approaches
\cite{gurkaynak,Lux-Sornette}, due to (i) the fact that the fundamental value
is in general poorly constrained and (ii) the difficulty in distinguishing
between an exponentially growing fundamental price and exponentially growing
bubble price. As we have already described, the JLS model defines a bubble in
terms of faster-than-exponential growth \cite{Johansen-Sornette}. Thus, the
main difference with standard bubble models is that the underlying price
process is considered to be intrinsically transient due to positive feedback
mechanisms that create an unsustainable regime. See for instance
Ref.\cite{Jiangetal09} where this is made as clear as possible.

\subsection{Which kind of bubbles can be detected by the JLS model?}

In page 4 of Ref.\cite{bree1}, three claims are outlined. One of them
states that: ``Financial crashes are preceded by bubbles with
fluctuations. Both the bubble and the crash can be captured by the
LPPL when specific bounds are imposed on the critical  parameters
$\beta$ and $\omega$'', where $\beta$ is presented as $m$ in this
paper.

For further clarification, this above claim is not entirely correct because
crashes can be endogenous or exogenous. The JLS model is suitable only for
endogenous crashes! Or more precisely, the JLS model is for bubbles, not for
crashes. Endogenous crashes are preceded by the bubbles that are generated by
positive feedback mechanisms of which imitation and herding of the noise
traders are probably the dominant ones among the many positive feedback
mechanisms inherent to financial system. Johansen and Sornette
\cite{Johansen-Sornette} identify 49 outliers in the distribution of
financial drawdowns, of which 25 can be classified as
endogenous crashes preceded by speculative bubbles, 22 as attributable to
exogenous events and fundamentally unpredictable and 2 as associated with the
Japanese anti-bubble \cite{JohansenSorJapan99}. Restricting to the world market indices,
they find 31 outliers, of which 19 are endogenous, 10 are exogenous and 2 are associated
with the Japanese anti-bubble. Although the endogenous outliers are more
frequent than the exogenous ones, the exogenous outliers still constitute a
quite large portion. This should not be too surprising as financial markets
are still impacted by news of all types that lead investors to reassess
their expectations of future risk-adjusted returns and therefore to rebalance their portfolio
allocations. In fact, according to the efficient market hypothesis, all crashes
should be exogenous, i.e. driven by external unanticipated news. The surprising
finding of this last decade, notably based on the JLS-LPPL model, has been
to demonstrate that about two-thirds of crashes are of an endogenous nature,
thus changing completely the paradigm of standard financial thinking.
It is thus important to stress that, while relevant for about
two-thirds of the observed crashes, the JLS model cannot capture all of the crashes
in the market. Only endogenous crashes that are preceded by bubbles can be
described by the JLS model. And when we write ``describe'', we mean that
the probability of crashes can be estimated, but the specific unfolding of
each crash and the post-crash dynamics is beyond the premise of the JLS model.

\section{Fitting Problems Concerning the JLS Model}
\label{sec:fitting}

\subsection{Extensions of the JLS model and their calibration}

The form of the JLS model represented by expression (\ref{eq:lppl}) is called the
first-order LPPL Landau JLS model. Extensions have been proposed, essentially
amounting to choose alternative forms of the crash hazard rate $h(t)$ that
replace expression (\ref{eq:hazard}). Let us mention the so-called second-order
and third-order LPPL Landau models
\cite{JohansenSornette01,JohansenSorJapan99,JohansenSorJapan00,SorZhoudeeper02,ZhouSorstab05},
the Weierstrass-type LPPL model \cite{GluzmanSornette02b,zs03anti}, the JLS
model extended with second-order and third-order harmonics
 \cite{ZhouSorturb02,SorZhoufuel04,ZhouSorrealest08,ZhouSorrSA09}
 and the JLS-factor model in which the LPPL bubble component is augmented
 by other financial risks factors \cite{ZhouSorFactor06,YanWoodSordiv12}.
 We should also mention that a non-parametric estimation of the log-periodic power law
 structure has been developed to complement the above parametric calibrations \cite{ZhouSornonpara03}.
 These extensions are warranted by the fact that the positive feedback
 mechanisms together with the presence of the symmetry of discrete
 scale invariance can be embodied in a general renormalization group equation
 \cite{GluzmanSornette02b}, whose general solution is the generalized
 Weierstrass LPPL model. Then, the first-order LPPL Landau JLS model
 can be considered to be just the first term in a general log-periodic
 Fourier series expansion of the general solution. Therefore, further
 away from the critical time $t_c$, corrections from the first-order
 expression can be expected to be relevant, depending on the context.
 In addition, nonlinear extensions to the renormalization group are embodied
 partially in the second-order and third-order LPPL Landau models,
 which extend the time domain over which the model can be calibrated
 to the empirical data \cite{SornetteJohansen97}.

Sornette and Johansen \cite{SornetteJohansen97} discussed the difference
between the fitting results obtained using the first order and the second order
LPPL Landau-type JLS models. They used daily prices of the S\&P 500 index from
1980 to 1987. The results show that the fitting result of the second order form
is much better than the first order form, as based on the measure of residual
sum of squares. A standard Wilks test of nested hypotheses confirms the fact
that the second-order form provides a statistically significant improvement
over the first-order form (recall that the first-order LPPL Landau formula is
recovered as a special case of the second-order LPPL Landau formula, hence the
first model is nested within the second model).

\begin{figure}[htbp]
\centering
\includegraphics[width=\textwidth]{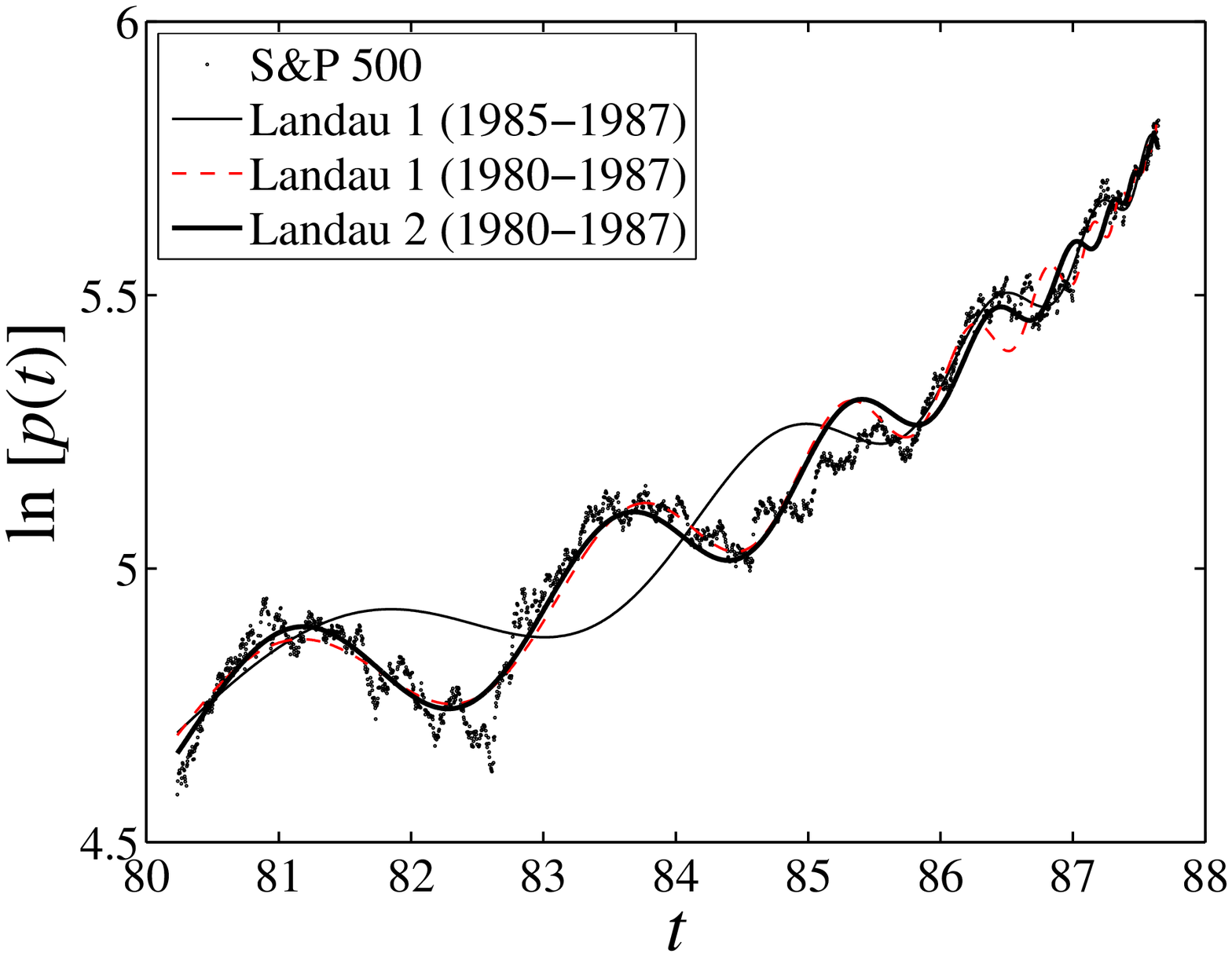}
\caption{Time dependence of the logarithm of the New York stock exchange index
S\&P 500 from January 1980 to September 1987 and the best fits by the first and
the second order LPPL Landau models. The crash of October 14, 1987 corresponds
to 1987.78 decimal years. The thin continuous line represents the best fit with the
first-order LPPL Landau model in the interval from July 1985 to the end of
September 1987 and is shown on the whole time span since January 1980.
The thin dashed line represents the best fit with the
first-order LPPL Landau model in the interval from from January 1980
to September 1987.  The thick line is the fit by the second-order LPPL Landau model
in the interval from January 1980
to September 1987. (Reproduced from \cite{SornetteJohansen97}) and extended).}
\label{fg:landaus}
\end{figure}

We reproduce the fitting
results from \cite{SornetteJohansen97} in Fig. \ref{fg:landaus} to give an
intuition on the difference between the first order and second order LPPL
Landau fits. When fitting the  first-order LPPL Landau formula
over the time period from January 1980 to September 1987 (dashed thin line),
we observe that the first-order LPPL Landau formula
accounts reasonably well for the price dynamics from 1980 to
1985 but then becomes completely out of phase in the remaining
two years before the crash. When fitting the first-order LPPL Landau formula
in the interval from July 1985 to the end of
September 1987 (continuous thin line) as done initially in Ref.\cite{SornetteJohansen97},
we observe that the first-order LPPL Landau formula accounts
reasonably well for the data from mid-1985 to the peak in October 1987
but is completely out of phase with the price in the earlier time period
from 1980 to 1985. In contrast, the second-order LPPL Landau formula (continuous thick line)
provides a good fit over the whole period from January 1980 to September 1987.
These results help explain why the results quoted
by Bree et al. \cite{bree2} for time windows of 834 trading days may be
questionable.

Notwithstanding the improvement provided by the second-order LPPL Landau model
for large time windows, it is sufficient in many cases to use the first-order
version just to get a diagnostic of the presence of a bubble. This is true even
when the time window is larger than 2-3 years. For instance, the first-order
LPPL Landau model was implemented within a pattern recognition method
\cite{SorZhouforecast06,rebound,YanRebWooSor} with time windows of up to 1500
days. The key to the reported performance in forecasting is the combination of bubble
diagnostics at multiple time scales, with common model parameters associated
with robustness \cite{SorZhouforecast06,rebound,YanRebWooSor}.

\subsection{Selection of the start of the time window}
\label{sec:t1selection}

A common question arising in fitting the JLS model is to decide which date
$t_1$ should be selected as the beginning of the fitting time window. Bree and
Joseph \cite{bree1} use a more consistent approach in defining the start time of the bubbles they analyze
than the papers published from 1998 to 2000 that they compare with.
It is worthwhile to mention that there are other procedures implemented systematically in
the more recent Refs. \cite{Jiangetal09,rebound,phpro,pp1,BFE-FCO09,fg},
in which multiple starting dates $t_1$'s are selected to make the prediction more
statistically robust. The findings of these articles is that
focusing on a single $t_1$ --- corresponding to a single time window in which to perform the fit --- may be unreliable and
an ensemble of fits with different $t_1$ is recommended. Indeed, this
reflects the fact that the start of a bubble is an elusive concept, as the early
deviation of the observed price from its fundamental value is small in the
first month and even years of the bubble, so the definition of the
start date is somewhat fuzzy and degenerate.

\subsection{Should price or log-price be fitted?}

Sornette and Johansen \cite{js} argue that log-price should be used when the
amplitude of the expected crash is proportional to the price increase during
the bubble. This prescription has been followed in
many of the studies on the JLS model
\cite{Jiangetal09,Zhou-Sornette-2003a-PA,BFE-FCO09,BFE-FCO10,BFE-FCO11,YanRebWooSor,rebound,pp1}.
This is because (\ref{eq:lppl}) is derived from (\ref{eq:jlsdynamics}), which
assumes that the changing price $dp$ is proportional to the price $p$.
Therefore, this statement is in accordance with Bree et al.'s definition of a
crash (25\% drop in price) in \cite{bree1,bree2}. Hence, it seems that the
approach by Bree et al.~to compare the results of the fits when using the price
(and not the log-price) is inconsistent.

One can also investigate the possibility that price changes may not be
proportional to price. If this is the case, use of the real price is warranted
according to the arguments put forward by Sornette and Johansen \cite{js}. In
practice, it is useful to try both fitting procedures with prices and
log-prices and compare their relative merits. But one should be cautious
because the fits using prices (and not log-prices) involve data values that may
change over several orders of magnitude over the time window of interest. As a
consequence, the standard least square fits is not suitable anymore. Instead, a
normalized least square minimization is recommended so that each data point of
the time series roughly contributes equally to the mean-square root diagnostic.
This approach has been implemented recently in Ref.~\cite{pp1}.

\subsection{Sloppiness of the JLS models and search algorithm}
\label{sec:search}

Bree and Joseph \cite{bree1} and Bree et al.~\cite{bree2} claim that the
concept of sloppiness and its consequence should be considered in fitting the
JLS model to empirical data with the Levenberg-Marquart algorithm. They
challenge the relevance of the obtained fits. We agree that sloppiness is real,
but is not an issue with an appropriate methodology addressing it, and
in view of the fact that the probabilistic predictions are in general consistent.

Specifically, a correct fitting procedure should include the combination of
the Levenberg-Marquart algorithm \cite{kl} \emph{and} a preliminary taboo
search \cite{ck} or other meta-heuristics such as the genetic algorithm and
simulated annealing algorithm. This should occur together with the slaving of
the linear parameters to the nonlinear ones in order to reduce
the nonlinear searching space from 7 dimensions to 4 (and to 3 in the recently
novel procedure of Filimonov and Sornette \cite{filisor11}). The taboo search
is a very good algorithm that provides a robust preliminary systematic
exploration of the space of solutions, which prevents the Levenberg-Marquart
algorithm later on being stuck in special regions of the space of solutions.
Also in a standard fitting procedure, the many results that may be obtained
from the taboo search (i.e. results associated with different parts of the
searching space) should be kept.

Taking into account the two points mentioned above, the quality of the fits
with the JLS model is furthermore checked by the visualization of the
fits and the original data \cite{oil,Jiangetal09,ZhouSorrSA09,repo,pp1}. In
contrast, it is obvious that fits using only the Levenberg-Marquart algorithm
without a reasonable starting value and sufficient preliminary exploration of
the space of solutions will produce spurious results, with the algorithm stuck
in a local minimum at the boundary of the search space.
%A typical example of such fitting failure is shown in Ref.~\cite{bree2}, where all the fitted $m$ values are either very close to 0 or close to 1 and almost all the fitted $t_c$ and $\omega$ values are very close to 0.

Bree and Joseph \cite{bree1} provide a sensitivity analysis of the root mean
square error (RMSE), in which one parameter is scanned while the others remain
fixed. The problem is that, because of the nonlinearity of expression
(\ref{eq:lppl}), it is not obvious that the results of such a scan can be
trusted.  That is, if local minima in, say, $\omega$ are found while the other
parameters are kept fixed, do the same minima appear when one or more of the
other parameters are changed to different values?  In other words, is the
multi-dimensional parameter landscape around these minima smooth?  The answer
to this question is more important than showing the sensitivity of 2
dimensional subspaces, as in \cite{bree1}. In practice, answering this question
on the smoothness of the multi-dimensional parameter landscape is difficult.
Filimonov and Sornette \cite{filisor11} have documented that the standard
slaving of three linear parameters ($A, B$ and $C$ in expression
(\ref{eq:lppl})) to the four remaining nonlinear parameters results in a
corrugated fitness landscape that requires a meta-heuristic (such as the taboo
search). This approach simultaneously changes all parameters in order to find acceptable
minima as starting points for the Levenberg-Marquart algorithm. Yes, this
approach does not guarantee finding \emph{the} absolute minimum but it does
provide \emph{an ensemble of acceptable local minima}. This ensemble approach
is more robust than searching in vain for a single global minimum.

\subsection{Performance of the recommended fitting method on synthetic data}

It is an essential building stone of any fitting procedure that it should be
tested on synthetic data. Indeed, in any calibration exercise, one faces
simultaneously two unknowns: (i) the performance, reliability and robustness of
the calibration procedure and (ii) the time series under study from which one
hopes to extract meaningful information. As a first step in the analysis, we
explore how the fitting procedure performs when the underlying data-generating
process is known. Johansen et al. \cite{jls,jsl} set the stage by developing
comparative tests on synthetic time series generated by the GARCH model. We
also attract the attention to the fact that one of the most extensive set of
synthetic tests concerning the possible existence of spurious log-periodicity
is found in reference \cite{Huangetalarti00}. Zhou and Sornette \cite{zs2002}
presented a systematic study of the confidence levels for log-periodicity only,
using synthetic time series with many different types of noises, including
noises whose amplitudes are distributed according to power law distributions
with different exponents and long-memory modeled by fractional Brownian noises
with various Hurst exponents spanning the full range from anti-persistent ($0 <
H < 1/2$) to persistent ($1/2 < H < 1$).

We now show that the current fitting methods estimate the parameters of the JLS
model within a reasonable range of uncertainty. For this, a
reference log-periodic power law (LPPL) time series of duration equal to 240
days is generated for a typical set of parameters, shown in
Table~\ref{tb:origin}.  This series corresponds to a value of the critical time
$t_c$ equal to 300 (days).  The choice of 240 days for the time window size is
motivated by the typical length for the generation of bubbles found in various
case studies in the literature. The choice of $\omega =10$ is in the range
often found in empirical studies. We note that there are actually
several ranges for $\omega$ corresponding to respectively the first, second
and third harmonics of the log-periodicity, as documented for instance
in Refs.\cite{ZhouSorturb02,SorZhoudeeper02,ZhouSornonpara03,Zhou-Sornette-2003a-PA,ZhouSorchina04,zhoudunbar05}.

\begin{table}
\begin{tabular}{|r|r|r|r|}
\hline & Reference & Mean (std) of Gaussian & Mean (std) of Student's t
\\\hline $t_c$&300& 296.07 (20.44)&295.15 (20.81)\\\hline $m$&0.7&0.74 (0.15) &
0.72 (0.18)\\\hline $\omega$&10&9.75 (1.43)&9.71 (1.47)\\\hline
\end{tabular}
\caption{\label{tb:origin} The parameter values used to generate the synthetic
data are shown in the second column ``Reference''. The mean and standard
deviation values of the parameters obtained by fitting the JLS model to the
synthetic LPPL time series decorated by the two types of noise discussed in the
text are given in the last two columns. These numbers are estimated from 200
statistical realizations of the noise, and each realization is characterized by
ten different best fits with the Levenberg-Marquart algorithm, leading to a
total of 2000 estimated parameters. The other parameters used to generate the
synthetic LPPL are $\phi = 1, A = 10, B = -0.1, C = 0.02$.}
\end{table}

\begin{figure}[htbp]
\centering
\includegraphics[width=\textwidth]{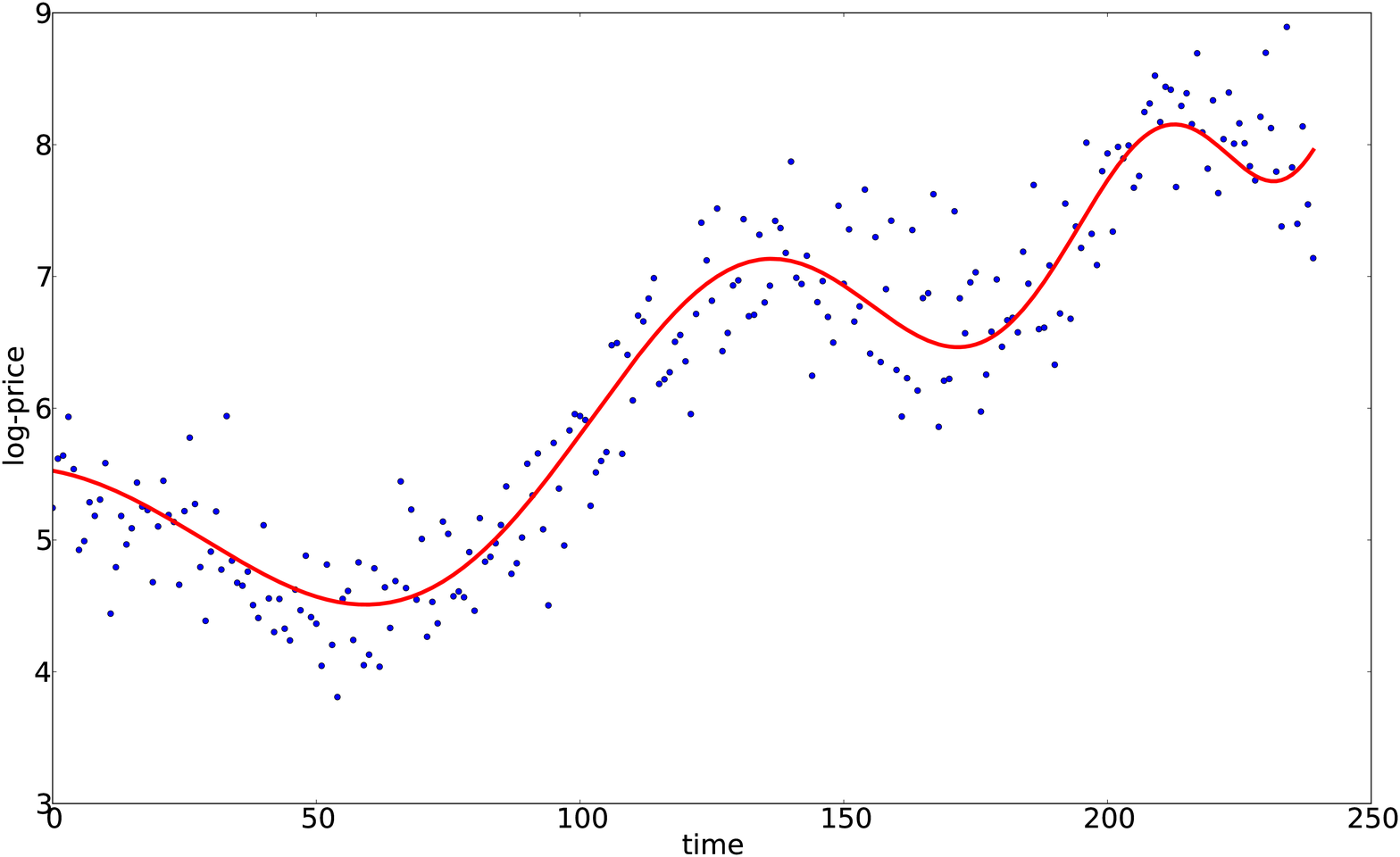}
\includegraphics[width=\textwidth]{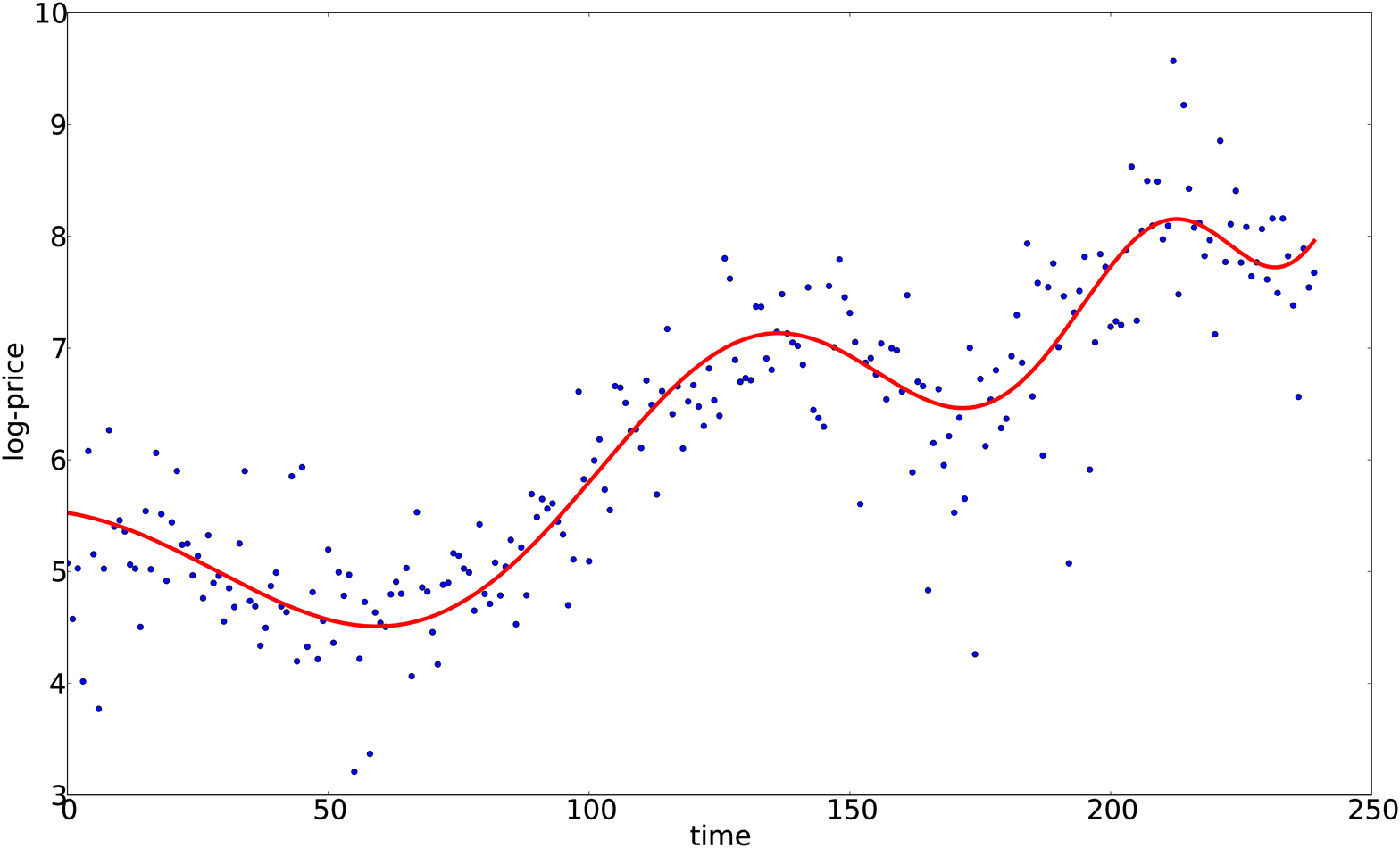}
\caption{Synthetic data examples with noise characterized by zero mean and large standard deviation
(5\% of the largest log-price among 240 reference points). Upper panel: the
synthetic data with Gaussian noise. Lower panel: the synthetic data with noise
generated with a Student t distribution with four degrees of freedom. The red
solid line shows the reference LPPL time series.} \label{fg:syntheticdata}
\end{figure}

The synthetic data is generated by combining the LPPL time series with noise.
Two kinds of noise are considered: Gaussian noise and noise generated with a
Student t distribution with four degrees of freedom (which exhibits a tail
similar to that often reported in the literature for the distribution of
financial returns). For both types of noise, the mean value is zero and the
standard deviation is set to be 5\% of the largest log-price among the 240
observations in the reference series. The standard deviation is chosen quite
high in order to offer a stringent test of the efficiency of the current fitting
method. Synthetic samples obtained with both types of noise along with the
reference time series are shown in Fig.~\ref{fg:syntheticdata}.

\begin{figure}[htbp]
\centering
\includegraphics[width=\textwidth]{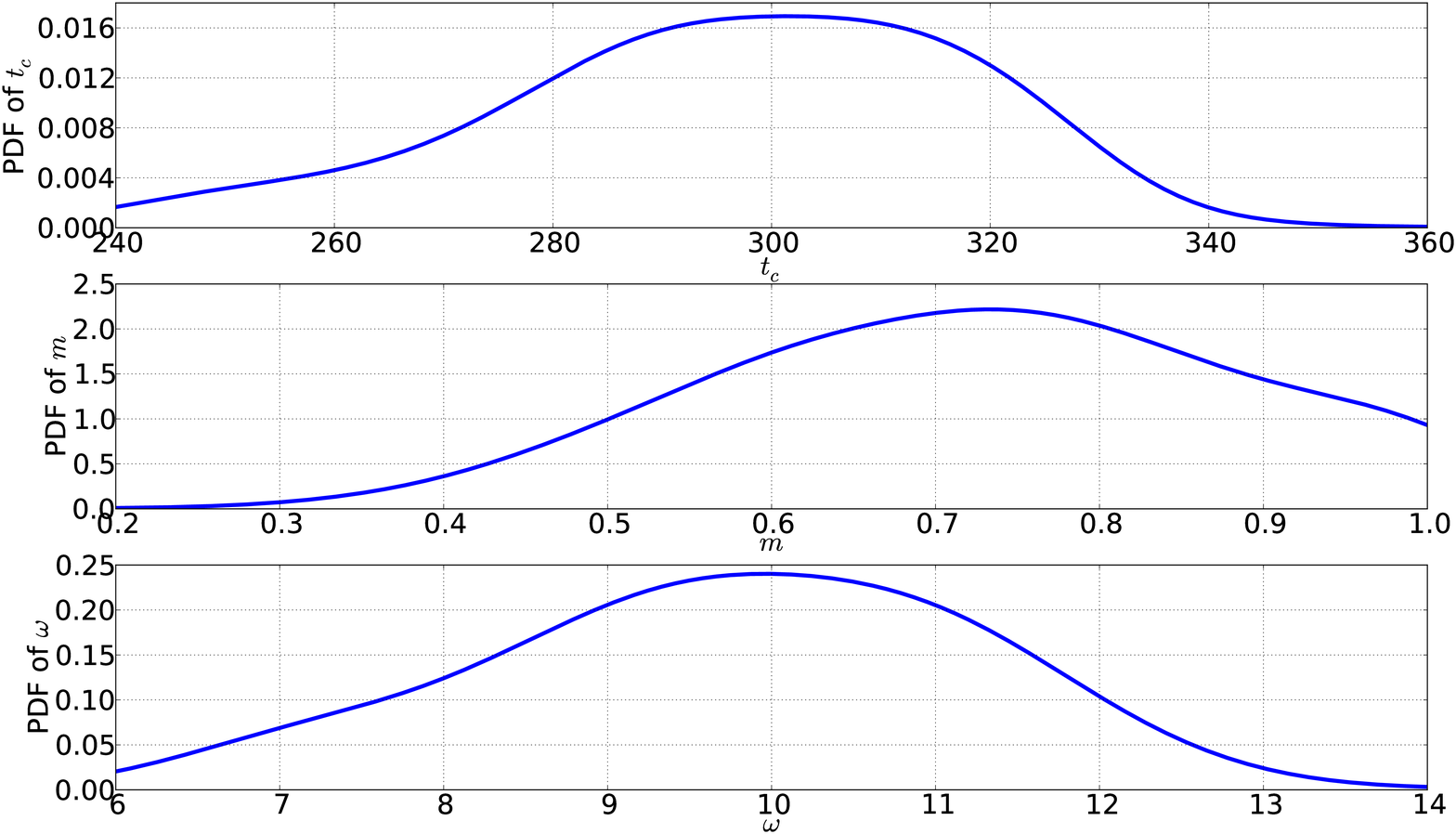}
\includegraphics[width=\textwidth]{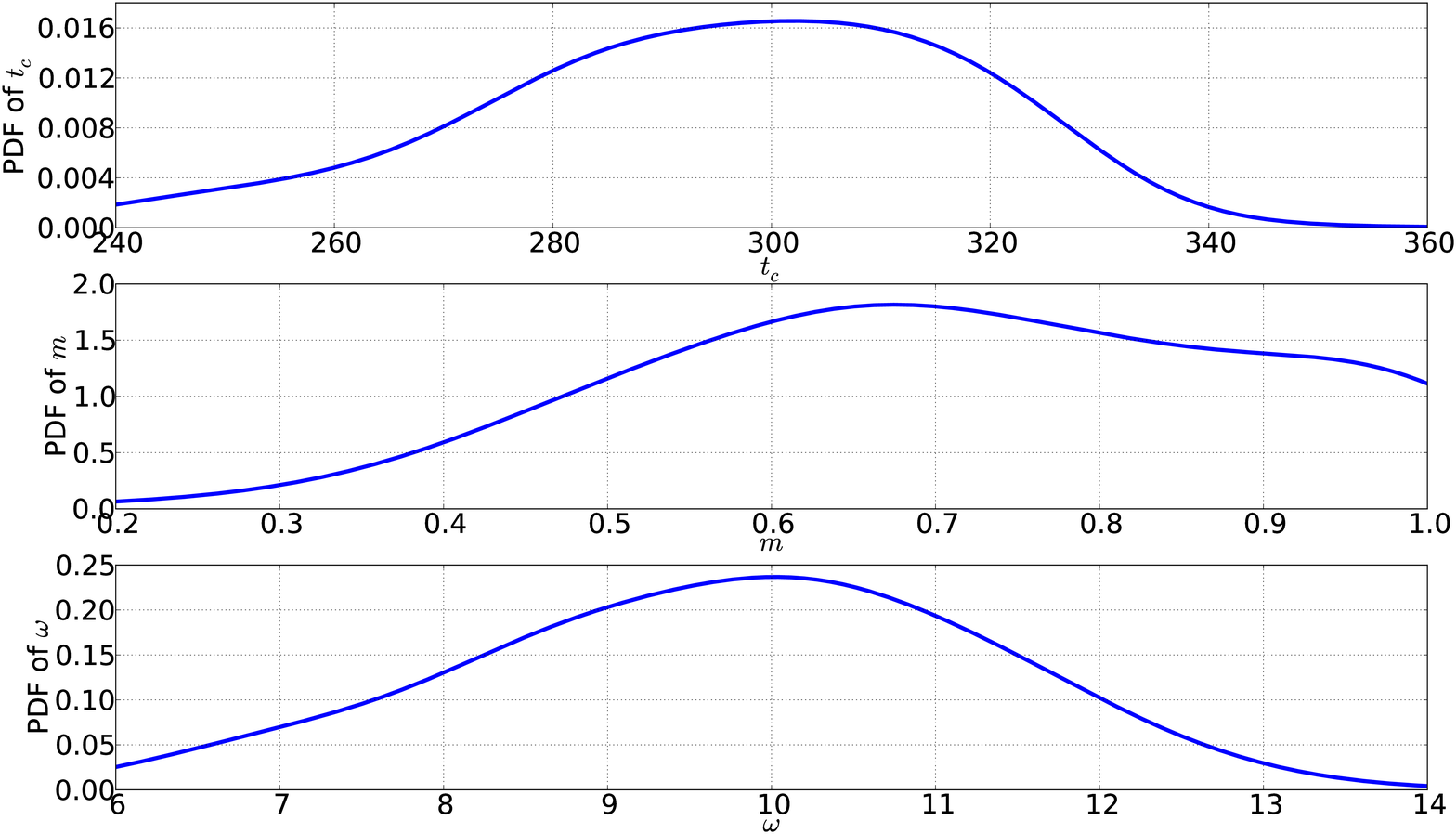}
\caption{The sampling distributions of $t_c$, $m$ and $\omega$ obtained by a
non-parametric kernel method applied to the parameter values determined by
analyzing 200 synthetic time series, each of which being characterized by its
ten best fits with the Levenberg-Marquart algorithm, leading to a total of 2000
estimated parameters. Upper panel: the synthetic data with Gaussian
noise. Lower panel: the synthetic data with noise generated with a Student t
distribution with four degrees of freedom.} \label{fg:syntheticpdf}
\end{figure}

For each type of noise, 200 synthetic time series are generated. We fit each
series with the JLS model (\ref{eq:lppl}) and keep the ten best fits for each
one. Recall that our stochastic fitting method produces multiple `good' fits
instead of the `best' fit, which, in practice, is difficult, if not impossible,
to find. In the new procedure recently developed by Filimonov and Sornette
\cite{filisor11}, the `best' fit can be found in most cases that are qualified
to be in a bubble regime. However, we still use the standard heuristic
procedure in the present paper, which predates that of
Filimonov and Sornette  \cite{filisor11}. The best ten selections result in 2000 sets of
estimated parameters for each type of noise. The sampling distributions of
$t_c$, $m$ and $\omega$ for the two types of noise are calculated by a
non-parametric method (adaptive kernel technique). The results are shown is
Fig.~\ref{fg:syntheticpdf}.

The mean and standard deviation of these parameters are shown in
Table~\ref{tb:origin} alongside the original numbers used to
generate the true LPPL function ($t_c = 300, m = 0.7, \omega = 10$) without noise.
This test on synthetic data demonstrates that the fitting method
combining the meta-heuristic Taboo search with the  Levenberg-Marquart algorithm
is satisfactory. We observe negligible biases, especially for the crucial
critical time parameter $t_c$. The standard deviation for $t_c$ of about $20$ days
is three times smaller than the 60 days separating the last observation
(day $240$) of the time series and the true critical time occurring at the $300$-th day,
showing that the calibration of a time series exhibiting LPPL structure, even
with very large statistical noise, can provide significant skills in forecasting
the critical time $t_c$.

\section{Probabilistic Forecast}
\label{sec:predict}

From a practical risk management view point, one of the benefits obtained from
the calibration of the JLS model to financial time series is the estimation of
the most probable time of the end of the bubble $t_c$, which can take the form
of a crash, but can also be a smooth transition to a new market regime.

As we mentioned before, a distribution of $t_c$ is obtained for a single bubble
period, associated with the set of fitted time windows (see Section
\ref{sec:t1selection}) and the recording of multiple locally optimal fits from
the stochastic taboo search (see Section \ref{sec:search}). Recall that the
output of the meta-heuristic is used as the initial guess required by the
Levenberg-Marquart algorithm. As demonstrated in the previous subsection, the
estimation of the distribution of the most probable time $t_c$ for the end of
the bubble is generated by a reliable non-parametric method \cite{liracine}.

Bree et al. \cite{bree2} make the interesting remark that the estimation of the
probability density of $t_c$ might be improved by augmenting the analysis of
the original time series with that of many replicas. These replicas of the
initial time series can be obtained for instance by using a LPPL function
obtained from the first calibration on the original time series and adding to it
noise generated by an AR(1) process. This methodology provides a measure of
robustness of the whole estimation exercise. The choice of an AR(1) process for
the noise is supported by the evidence provided in Refs. \cite{lin,gazola} that
the residuals of the calibration of the JLS model to a bubble price time series
can be reasonably described by an AR(1) process.

However this is only one among
several possibilities. Another method, which we have implemented in our group
for quite some time and now use systematically, is to generate bootstraps in
which the residuals of the first calibration on the original time series are
used to seed as many synthetic time series as needed, using reshuffled blocks
of residuals of different durations. For instance, reshuffling residuals in
blocks of 25 days ensures that the dependence structure between the residuals
is identical in the synthetic time series as in the original one up to a month
time scale.  Note that this bootstrap method does not assume Gaussian residuals
in contrast with the AR(1) noise generation model. It captures also arguably
better the dependence structure of the genuine residuals than the linear
correlation embedded in the AR(1) model.

\section{Conclusion}
\label{sec:conclusion}

We have discussed the present theoretical status and some calibration issues
concerning the Johansen-Ledoit-Sornette (JLS) model of rational expectation
bubbles with finite-time singular crash hazard rates. We have provided a guide
to the advances that have punctuated the development of tests of the JLS model
performed on a variety of financial markets during the last decade. We can say
that the development of new versions and of methodological improvements have
paralleled the occurrence of several major market crises, which have served as
inspirations and catalyzers of the research. We believe that the field of
financial bubble diagnostic \cite{BFE-FCO09,BFE-FCO10,BFE-FCO11,Sorwoodtokyo} is
progressively maturing and we foresee in the near future that it could become
operational to help decision makers alleviate the consequences of excess
leverage leading to severe market dysfunctions.

\vskip 0.5cm
{\bf Acknowledgement}: Wei-Xing Zhou acknowledges financial support from
 the National Natural Science Foundation of China
(11075054), and the Fundamental Research Funds for the Central
Universities. This work was partially supported by ETH Research Grant ETH-31 10-3
``Testing the predictability of financial bubbles and of systemic instabilities''.

\clearpage

\end{document}